# Emergent phase in short-periodic rare-earth nickelate superlattices

*S. Middey,\* Ranjan Kumar Patel, D. Meyers, Xiaoran Liu, M. Kareev, P. Shafer, J.-W. Kim, P. J. Ryan, and J. Chakhalian*

Dr. S. Middey, Ranjan Kumar Patel
Department of Physics, Indian Institute of Science, Bengaluru 560012, India
Email: smiddey@iisc.ac.in

Dr. D. Meyers
Department of Physics, Oklahoma State University, Stillwater, Oklahoma 74078, USA

Dr. Xiaoran Liu, Dr. M. Kareev, Prof. Dr. J. Chakhalian
Department of Physics and Astronomy, Rutgers University, Piscataway, New Jersey 08854, USA

Dr. P. Shafer
Advanced Light Source, Lawrence Berkeley National Laboratory, Berkeley, California 94720, USA

Dr. J.-W. Kim, Dr. P. J. Ryan,
Advanced Photon Source, Argonne National Laboratory, Argonne, Illinois 60439, USA

Dr. P. J. Ryan
School of Physical Sciences, Dublin City University, Dublin 11, Ireland

Heterostructure engineering provides an efficient way to obtain several unconventional phases of $LaNiO_3$, which is otherwise paramagnetic, metallic in bulk form. In this work, a new class of short periodic superlattices, consisting of $LaNiO_3$ and $EuNiO_3$ have been grown by pulsed laser interval deposition to investigate the effect of structural symmetry mismatch on the electronic and magnetic behaviors. Synchrotron based soft and hard X-ray resonant scattering experiments have found that these heterostructures undergo simultaneous electronic and magnetic transitions. Most importantly, $LaNiO_3$ within these artificial structures exhibits a new antiferromagnetic, charge ordered insulating phase. This work demonstrates that emergent properties can be obtained by engineering structural symmetry mismatch across a heterointerface.

# 1. Introduction

The quest to elucidate the origin of simultaneous metal-insulator transition (MIT), charge ordering (CO) and structural symmetry change in rare earth nickelate (*RE*NiO$_3$) series have drawn significant attentions in recent times.[1,2] Another simultaneous magnetic transition in the case of NdNiO$_3$ and PrNiO$_3$ adds further complexity to the problem. In contrast, LaNiO$_3$ (LNO) is the only member of the series which has rhombohedral structure and remains metallic down to the lowest temperature.[3,4] Moreover, a recent report of paramagnetic to *E'*-type antiferromagnetic (*E'*-AFM) transition in the metallic LNO single crystal has prompted active debates and sparked reinvestigation of the magnetic nature of the ground state of LNO.[5-7] To add further complexity, these compounds are also characterized as the negative effective charge transfer ($\Delta'$) systems.[8-12] Interestingly, the appearance of an antiferromagnetic metallic phase was predicted earlier by the model Hamiltonian calculations for negative $\Delta'$ system.[13] Additionally, such an antiferromagnetic metallic system has excellent potential for spintronics applications.[14] These findings motivate investigation of how to create an antiferromagnetic, charge ordered ground state in LNO. Heterostructuring of LNO has been shown to enable several unconventional behaviors (see Refs.1,2 and references therein). For example, unlike the negligible orbital polarization of bulk LNO, LaNiO$_3$/LaTiO$_3$ heterostructure exhibits strong orbital anisotropy.[15-17] Helical spin configurations were realized in LNO through interfacial charge transfer from La$_{0.67}$Sr$_{0.33}$MnO$_3$ layers.[18] Superconductivity was also reported in LNO/La$_{0.7}$Sr$_{0.3}$MnO$_3$ superlattice.[19] The reduction of LNO film thickness results in an insulating phase due to quantum confinement effect.[20-25] A ferromagnetic phase in epitaxial LNO film on LaAlO$_3$ (1 1 1) substrate was also demonstrated.[26] Surprisingly, magnetic ordering, analogous to *E'*-AFM phase and without any charge ordering was obtained in 2 uc of LNO layers sandwiched between the insulating spacer LaAlO$_3$.[23,27] Despite all these findings and somewhat surprisingly, long-range charge ordering (CO) has not been demonstrated thus far.

Apart from the above mentioned $RE$NiO$_3$/$AB$O$_3$ ($B\neq$Ni) heterostructures, epitaxial growth of two different members of the $RE$NiO$_3$ series in a layer-by-layer way offers another interesting route to achieve unconventional behavior.[28-30] For example, bulk EuNiO$_3$ (ENO) undergoes simultaneous electronic and structural transition at 410 K and a magnetic transition around $T_N \sim$ 200 K. Surprisingly, the artificial structure in the form of [1 uc ENO/1 uc LNO] (uc= unit cell in pseudocubic setting) superlattice (SL) grown on single crystalline NdGaO$_3$ exhibit all four simultaneous transitions, similar to bulk NdNiO$_3$ and PrNiO$_3$.[28] As illustrated in **Figure 1**(a), each Ni in the 1ENO/1LNO SL [marked as Ni(I)] has Eu on one side and La on the opposite side along [0 0 1]$_{pc}$, implying each layer is equivalent to an ordered analog of Eu$_{0.5}$La$_{0.5}$NiO$_3$ composition.[30] On the contrary, [m uc EuNiO$_3$/m uc LaNiO$_3$] (mENO/mLNO) superlattices with m >1 have additional bulk-like ENO and LNO unit cells, where each Ni has either Eu or La on both sides (respectively marked by Ni(II) and Ni(III) in Figure 1(b)). Since bulk ENO and LNO have a different octahedral rotational pattern $a^-a^-c^+$ vs. $a^-a^-a^-$, strong structural competition between ENO and LNO layers in these [mENO/mLNO] systems can lead to new electronic and magnetic phases.[31,32]

Here we report the development of a new LNO-based heterostructure and demonstrate emergence of a novel charge ordered antiferromagnetic insulating phase of LNO. To avoid complications caused by quantum confinement and/or interfacial charge transfer, we have judiciously built a series of [$m$ EuNiO$_3$/$m$ LaNiO$_3$] X N superlattices (here N denotes the number of repeats) using pulsed laser deposition.[28-30] While the long-periodic superlattice ($m$=8) shows entirely metallic behavior, the short-periodic superlattices undergo simultaneous metal to insulator and paramagnetic to antiferromagnetic transitions. Resonant hard-X-ray scattering experiments have revealed the signature of monoclinic symmetry and Ni resonance features in the metallic phase of the $m$=5 film. The Ni resonance further increases after entering into the insulating phase, signifying the development of charge ordering throughout the whole superlattice. Magneto-transport measurements show a sign change in the Hall coefficient $R_H$ on crossing the transition temperature in all cases, which can be explained as a consequence of Fermi surface reconstruction. The combination of advanced experimental probes has confirmed the successful creation of a new antiferromagnetic and charge ordered insulating phase in LNO, which is otherwise a paramagnetic metal in the bulk form.

## 2. Results and Discussions

$m$ENO/$m$LNO superlattices were grown on single crystalline NdGaO$_3$ (1 1 0)$_{or}$ [(0 0 1)$_{pc}$] (here or and pc denote orthorhombic and pseudocubic setting) substrates using a pulsed laser deposition system.[28-30] For each SL, the deposition on NGO substrate starts with an ENO layer and terminates with a LNO surface layer. The sample thickness (2$m$N) is 36 uc for SLs with $m$ = 1, 3 and 30 uc for $m$=5 sample. The effect of tensile strain in LNO and ENO ultra-thin films is compensated by octahedral rotations and breathing-mode distortions, resulting in a bulk-like net out-of-plane lattice constant.[32-34] The effect of epitaxial strain of these SLs have been investigated using synchrotron X-rays. Figure 1(c) shows representative diffraction scans along the (0 0 l)$_{pc}$ truncation rod. As seen, the diffraction pattern of each SL consists of a sharp substrate peak, a broad film peak (indicated by #) and a set of Kiessig fringes. The presence of the satellite peaks further testifies for the intended superlattice structure of each film. Surprisingly, the out-of-plane lattice constants of all of these SLs ($c_{pc}$ = 3.793 Å, 3.793 Å and 3.791 Å for $m$=1, 3, 5 sample, respectively) are very similar to the $c_{pc}$ (3.8 Å) of EuNiO$_3$ thin film on NGO substrate.[34] This observation clearly suggests that in these short-periodic superlattices, the LNO layers follow the structural response of the ENO layers.

The modification in electronic behavior as a function of $m$ has been investigated by $dc$ transport measurements. Using a parallel resistor model with bulk-like ENO, bulk-like LNO and the interfacial ELNO layer, the expected resistance of the SL can be expressed as

$$R_{SL} = R_{ENO}R_{LNO}R_{ELNO}/(R_{ENO}R_{LNO} + R_{ENO}R_{ELNO} + R_{LNO}R_{ELNO}) \quad (1)$$

Here, all NiO$_2$ layers in $m$ENO/$m$LNO SLs having EuO layers on both sides along [0 0 1]$_{pc}$

are named as bulk-like ENO throughout the paper. Similarly, bulk-like LNO layers refer to the $NiO_2$ layers with LaO layers on both sides and interfacial ELNO layer refers to $NiO_2$ layer, having EuO in one side and LaO in other side. As expected from the **equation** 1, the long-period SL ($m$=8) is completely metallic (see **Figure 2**(a)). In sharp contrast, all short periodic SLs exhibit first order MIT as a function of temperature. The transport data of $m$=1 SL also show the first order MIT of the interfacial ELNO layer (marked as Ni(I) in Fig. 1). Unexpectedly, a large change in the layer number (e.g. $m$=1 → 5) within each period does not result in any drastic change in the transition temperature (**Table**-1). The observation of a MIT in the $m$=5 film implies that bulk-like LNO layers marked by Ni(III) in Figure 1(b) also become insulating below $T_{MIT}$ ~ 120 K. Apart from the bulk LNO, the metallic phase of $RE$NiO$_3$ show distinct non-Fermi liquid (NFL) behavior with the NFL exponent tunable by hydrostatic pressure or by epitaxial strain.[29,35-37] To extract the value of the power exponent $n$, we plot log($dR_s/dT$) vs. log($T$) shown in Figure 2(b), which yields n = 1.05±0.05 - the value which is very close to $n$=1 for $T$ > 175 K for the $m$=5 sample. Similar linear $T$-dependent behavior is also observed for 1ENO/1LNO SL.[29] Interestingly, $n$ (=1.3±0.02) is very close to 4/3 for 3ENO/3LNO SL in the range of 150 K < T <220 K but switches to unity above 240 K. Such NFL behaviors implies that the electronic behavior of LNO in these artificial structures is very different compared to the bulk LNO.

Next, we discuss the magnetic properties of the SLs. The insulating phase of all $RE$NiO$_3$ based systems host $E'$-AFM phase, characterized by (1/2 0 1/2)$_{or}$ [≡ (1/4 1/4 1/4)$_{pc}$] magnetic wave vector.[38] A reliable estimate of the magnetic transition temperature ($T_N$) can be obtained from resistivity analysis.[39,40] The $d(\ln R_s)/d(1/T)$ vs. $T$ plot (Figure 2(c)) yields $T_N$ of 145 K, 115 K and 110 K for SL with $m$=1, 3, 5 SL, respectively. On the other hand, resonant soft X-ray diffraction (RSXD) measured with the photon energy tuned to Ni $L_3$ edge is a direct method for the investigation of this peculiar $E'$-AFM phase.[27,28,30,36,41-45] Simultaneous MIT and magnetic transition for $m$=1 SL has been confirmed by our earlier RSXD experiments.[28] As shown in Figure 2(d), a (1/4 1/4 1/4)$_{pc}$ diffraction peak is also observed at 50 K in RXD of $m$=5 SL, which vanishes gradually with increasing $T$. A plot of the area under this peak as a function of $T$ gives $T_N \sim 115 \pm 5$ K, which is very close to the value of $T_N$ obtained from our transport data shown in Figure 2(c). Thus, apart from the interfacial unit cells, bulk-like LNO layers also undergo simultaneous metal-insulator and paramagnetic-antiferromagnetic transitions in these short periodic superlattices.

The microscopic origin of the simultaneous electronic and magnetic transitions in $RE$NiO$_3$ can be accounted by the hole-Fermi surface nesting driven spin-density wave (SDW) transition.[27,42,46,47] Interestingly, previous measurements of Hall coefficient ($R_H$) reveal that $R_H$ is positive in the paramagnetic metallic (PM) phase and becomes negative below $T_N$.[40] Such observation can be explained considering nesting driven transition of a multiband system like $RE$NiO$_3$. $R_H$ dominated by the contribution from large hole pockets is positive in the PM phase. It becomes negative below the SDW transition as the hole pockets are removed by the nesting and the remaining small electron pockets contribute to $R_H$ in the antiferromagnetic insulating phase. **Figure 3**(a)-(b) show the variation of transverse resistance ($R_{xy}$) as a function of the applied magnetic field ($H$) at different temperatures (measured in cooling run) for the SLs with $m$=3 and $m$=5. As seen, from the positive slope of the curves in Figure 3(a) $R_H$ (= $t dR_{xy}/dH$, $t$=film thickness) is hole-like above 100 K for the $m$=3 sample and then becomes electron-like. For the $m$=5 film, $R_H$ is also hole-like down to 120 K and is completely electron-like at 80 K. In the intermediate temperature range around 100 K, $R_{xy}$ shows non-linear behavior with $H$ and switches from electron- to hole-like behavior with increasing $H$. Such small value of $R_{xy}$ indicates that the difference in volume between electron and hole pockets is very small at that temperature. Similar switching of $R_H$ was also observed in 1ENO/1LNO SL (not shown). The crossover of $R_H$ in the vicinity of $T_{MIT} \sim T_N$ strongly point towards a SDW driven origin of the antiferromagnetic insulating (AFI) phase of these ENO/LNO superlattices.

After confirming the simultaneous electronic and magnetic transitions in SLs, we investigate the issue of the highly debated charge ordering. Unlike the bulk samples, thin films can host an insulating phase without any CO.[28,42,45,48] The conventional picture of charge disproportionation (CD) on Ni ions ($d^7+d^7 \rightarrow d^{7+\delta}+d^{7-\delta}$)[42,49-53] is also challenged by the alternative mechanism of bond disproportion (BD) transition, specially proposed for the most distorted members of the series.[54-58] Such BD transition ($d^8\underline{L}+ d^8\underline{L} \rightarrow d^8+d^8\underline{L}^2$, $\underline{L}$ represents a hole on oxygen) results in a rocksalt-type pattern with alternating $NiO_6$ octahedra with $d^8$, having long-bond (LB) and $d^8\underline{L}^2$ having short-bond (SB). Most importantly, the results of X-ray absorption spectroscopy and X-ray scattering experiments, conventionally used to conclude CD on Ni sublattices can be also accounted by the BD transition.[12,50-52,58-62] However, a larger amount of BD can generate a significant amount of Ni $e_g$ charge disproportionation. For example, double cluster calculation with a BD of $\delta d \sim 0.2$Å by Green et al. has not only obtained unequal spin values (S ~1.0 and 0.1 on LB and SB sites, respectively),[60] but also a difference of ~ 0.3$e$ in the $e_g$ electronic occupancy between LB and SB sites. It is important to note that the amount of CD, obtained experimentally, is of the same order 2$\delta$~ 0.45$e$.[50] This implies BD and CD are cooperative in nature for bulk rare-earth nickelates. On the other hand, Lee et al. have shown that CD on Ni sublattices can be realized as a simple consequence of the SDW transition.[46,47]

In order to investigate BD vs. CD in the weakly insulating phase of the $m$=5 film, we have measured resonant X-ray scattering at Ni $K$ edge (1$s \rightarrow 4p$) around (0 1 1)$_{or}$ [≡(-1/2 1/2 1/2)$_{pc}$] and (1 0 1)$_{or}$ [≡ (1/2 1/2 1/2)$_{pc}$] reflections.[28,50-52,61] The details of the energy and momentum dependence of these diffraction peaks are given in Ref. 63. Here we note that (-1/2 1/2 1/2)$_{pc}$ peak is allowed for monoclinic and forbidden for orthorhombic symmetry. On the other hand, the (1/2 1/2 1/2)$_{pc}$ reflection is allowed in both monoclinic and orthorhombic phase; both reflections are forbidden for rhombohedral symmetry. As seen in **Figure** 4(a), $m$=5 sample exhibits a resonance feature for (-1/2 1/2 1/2)$_{pc}$ diffraction peak around 8.345 keV photon energy at 50 K. Generally, both the resonant intensity ($I_{res}$) and the off-resonant intensity ($I_{offres}$) decrease with the increase of $T$ as previously reported for the insulating phase of thick NdNiO$_3$ film and for 1ENO/1LNO SL.[28,50,61] However, $I_{offres}$ increases strongly up to 90 K for $m$=5 SL (Figure 4(b)), suggesting the existence of additional structural effects below 90 K. Interestingly, $I_{res}$ is also non-zero in the metallic phase and the line shape is very similar to the resonant line shape observed in insulating of EuNiO$_3$ thin film.[44] The resonance and off-resonance intensity for the (1/2 1/2 1/2)$_{pc}$ diffraction peak increases up-to 90 K and shows a marked jump across the insulator to metal transition (see Figure 4(d), (e)). The observation of both (-1/2 1/2 1/2)$_{pc}$ and (1/2 1/2 1/2)$_{pc}$ reflections and Ni resonance in the metallic phase of the SL clearly establish the presence of strong monoclinic distortion in the metallic phase of the $m$=5 SL. Moreover, the temperature dependence of the resonance enhancement factor [$\sqrt{I_{res}}$ -$\sqrt{I_{offres}}$] for (-1/2 1/2 1/2)$_{pc}$ diffraction peak (Figure 4(c)) demonstrates the onset of a charge ordering transition around 110 K.

From these experimental results, we suggest the following scenario for simultaneous metal-insulator, CO and magnetic transitions in these SLs. As shown in Figure 1(a), $m$=1 SL consists of only interfacial Ni's, having EuO on one side and LaO layers on the opposite side. The site-centred type SDW transition (Figure 4(f)) in all of these interfacial layers can give rise to the AFI phase with CO. Apart from the interfacial NiO$_2$ layers, $m$=5 SL also contain bulk-like ENO and bulk-like LNO layers. The interfacial layers and the bulk-like LNO layers remain metallic down to 120 K and the bulk-like ENO layers are charge ordered insulator with monoclinic symmetry, resulting in overall metallic behavior of the sample. In addition to the interfacial layers, bulk-like LNO layers also undergo SDW transition to create the insulating phase with antiferromagnetism and charge ordering (Figure 4(f)). Unlike the pure SDW phase with CO obtained in LaNiO$_3$/LaAlO$_3$ SL through dimensionality control,[23,27,61] LaNiO$_3$ layers in the present study undergo SDW transition in-spite of the presence of three dimensional NiO$_6$ octahedral networks, resulting in the emergent charge ordered, antiferromagnetic phase of LaNiO$_3$, which is unattainable in the bulk form.

## 3. Conclusion

To summarize, our detailed measurements including magneto-transport, hard X-ray resonant scattering and soft X-ray resonant magnetic scattering have found simultaneous metal-insulator, charge ordering and magnetic transitions in the [$m$ EuNiO$_3$/$m$ LaNiO$_3$] SLs. In addition, the Hall coefficient shows crossover from hole-like to electron-like behavior across the transitions. All of these experimental observations can be successfully accounted by the notion of Fermi surface nesting emerging in the bulk-like LNO layers and the interfacial layer. These findings highlight the capability of heterointerface engineering to tailor new electronic and magnetic phases of complex oxide materials.

## 4. Experimental Section

The layer by layer growth of each unit cell of EuNiO$_3$ and LaNiO$_3$ in PLD process was confirmed by in-situ RHEED (reflection high energy electron diffraction). The details of growth conditions can be found in Ref. [28-30]. Sample quality was checked by X-ray diffraction using laboratory-based X-ray diffractometer and also by a six-cycle diffractometer in 6-ID-B beamline of Advanced Photon Source (APS) in Argonne National Laboratory. Magneto-transport properties of these samples were measured in four probe Van Der Pauw geometry using a Quantum design PPMS (Physical Property Measurement System). The issue of charge ordering in these thin films were investigated by resonant hard X-ray scattering experiments on Ni K edge in 6-ID-B beam line of Advanced Photon Source. Resonant soft X-ray scattering experiments at the Ni $L_3$ edge were performed at beam line 4.0.2 of the Advanced Light Source (ALS) to probe the $E'$-antiferromagnetic order.

**Acknowledgements**

SM acknowledges DST Nanomission grant no. DST/NM/NS/2018/246 and SERB Early Career Research Award (ECR/2018/001512) for financial support. J.C. is supported by the Gordon and Betty Moore Foundation EPiQS Initiative through Grant No. GBMF4534. This research used resources of the Advanced Photon Source, a U.S. Department of Energy Office of Science User Facility operated by Argonne National Laboratory under Contract No. DE-AC02-06CH11357. This research used resources of the Advanced Light Source, which is a Department of Energy Office of Science User Facility under Contract No. DE-AC02-05CH11231.

[63]   We recap that the scattering intensity of (011)or and (101)or peaks are given by

$I_{011}(\mathbf{Q},E) \propto [A^2_{O,RE}(\mathbf{Q})+2A_{O,RE}(\mathbf{Q}).2(\Delta f^0_{Ni}(E)+\Delta f'_{Ni}(Q)+\Delta f''_{Ni}(E)) + 4(\Delta f^0_{Ni}(E)+\Delta f'_{Ni}(E)+\Delta f''_{Ni}(E))^2]$

$I_{101}(\mathbf{Q},E) \propto [B^2_{O,RE}(\mathbf{Q})-2B_{O,RE}(\mathbf{Q}).2(\Delta f^0_{Ni}(E)+\Delta f'_{Ni}(E)+\Delta f''_{Ni}(E))+4(\Delta f^0_{Ni}(E)+\Delta f'_{Ni}(E)+\Delta f''_{Ni}(E))^2]$

with  $\Delta f^0_{Ni}(E)= [f^0_{Ni1}(E)- f^0_{Ni2}(E)]$, $\Delta f'_{Ni}(E)= [f'_{Ni1}(E)- f'_{Ni2}(E)]$, $\Delta f''_{Ni}(E)= [f''_{Ni1}(E)- f''_{Ni2}(E)]$.[28] $A_{O,RE}(\mathbf{Q})$, $B_{O,RE}(\mathbf{Q})$ represent the energy-independent Thompson scattering (TS) terms for the RE and O sites and $f^0_{Ni}(E)$ corresponds TS terms for the Ni sites. $f'_{Ni}(E)$ and $f''_{Ni}(E)$ term,s responsible for resonance behaviors, are the real and imaginary energy-dispersive correction factors. Since $f'_{Ni}(E)$ and $f''_{Ni}(E)$ are negligible away from the resonance, the off-resonant intensity represents pure structural information. The energy-independent term $A_{O,RE}(\mathbf{Q})$, and $B_{O,RE}(\mathbf{Q})$ can also result in energy dependent feature at resonance energy due to their coupling with the energy dependent term $\Delta f'_{Ni}$, $\Delta f''_{Ni}$. The minus sign in the last equation implies that the resonance feature around 8.345 keV would appear as a dip for (1 0 1)or diffraction peak. We also note that the off-resonance intensity of (1 0 1)or will be stronger compared to the (0 1 1)or peak due to the strong structural contribution via $B_{O,RE}$ term in the last equation.

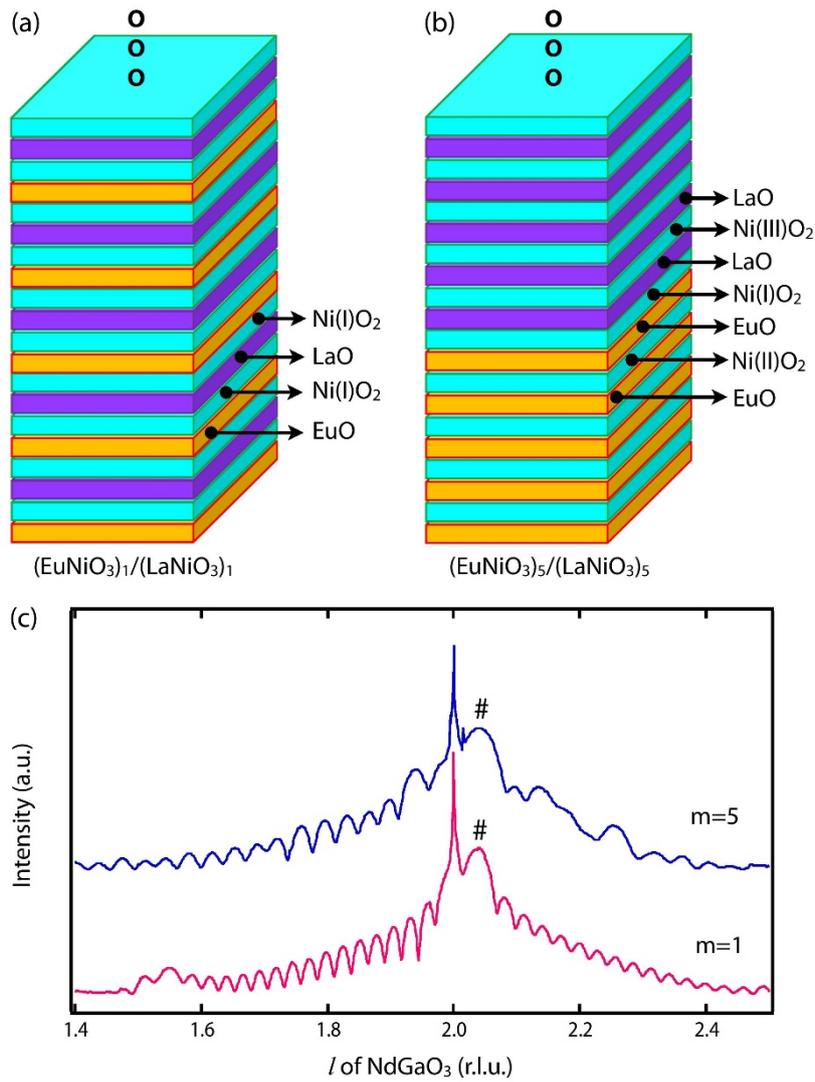

**Figure 1.** Schematics of atomic planes along $[0\ 0\ 1]_{pc}$ for (a) $m=1$ and (b) $m=5$ sample. Ni(I)O$_2$ layers have EuO on one side and LaO on the opposite side. Ni(II)O$_2$ [Ni(III)O$_2$] layers have neighboring EuO [LaO] layers on both sides. (c) $l$-scans through $(0\ 0\ 2)_{pc}$ truncation rod. # indicates the film peak. Data have been offset along the intensity axis for visual clarity.

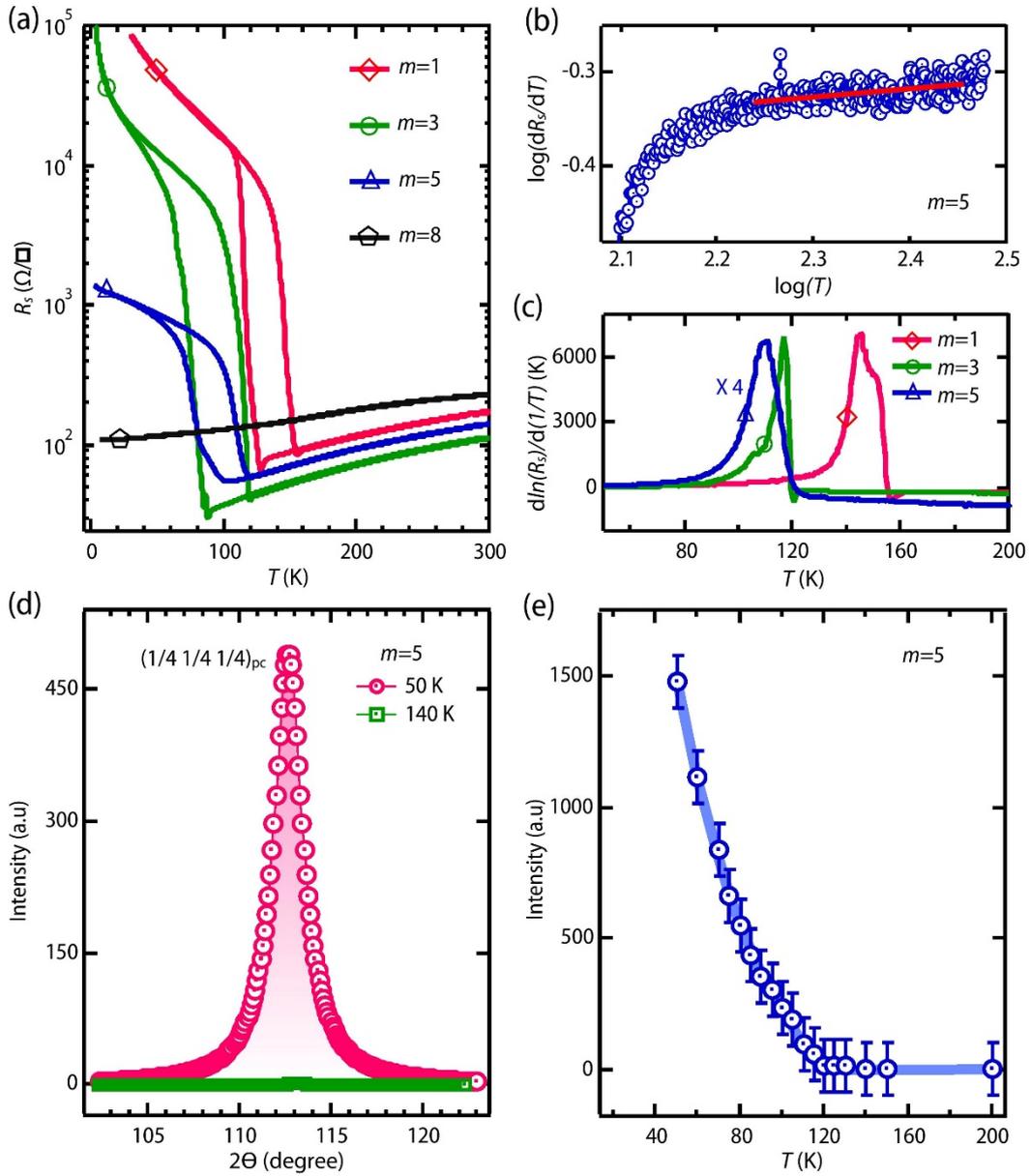

**Figure 2.** (a) Sheet resistance of $m$ENO/$m$LNO SLs. The data for $m$=1 sample has been adapted from Ref. 28. Reproduced with permission.[28] 2018, American Physical Society (b) log($dRs/dT$) vs. log($T$) plot for the $m$ =5 sample. The solid line represents a linear fitting of the experimental data. (c) Transport data of these SLs have been analyzed in terms of $d(\ln Rs)/d(1/T)$ vs $T$ plot to determine $T_N$. (d) Measured magnetic scattering for $m$ =5 sample at 50 K and 140 K. The integrated area of this peak has been plotted a function of $T$ in panel (e).

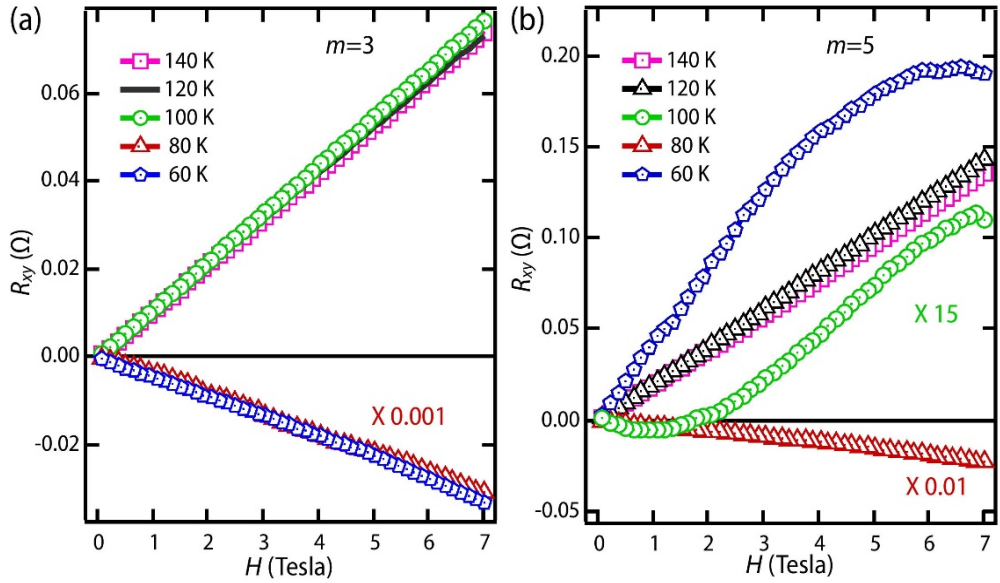

**Figure 3.** Transverse resistance $R_{xy}$ as a function of $H$ at different $T$ for (a) $m=3$ and (b) $m=5$ sample. Several data sets have been scaled to represent in same y-axis scale. $R_{xy}$ of select data sets are scaled and indicated by scaling factors.

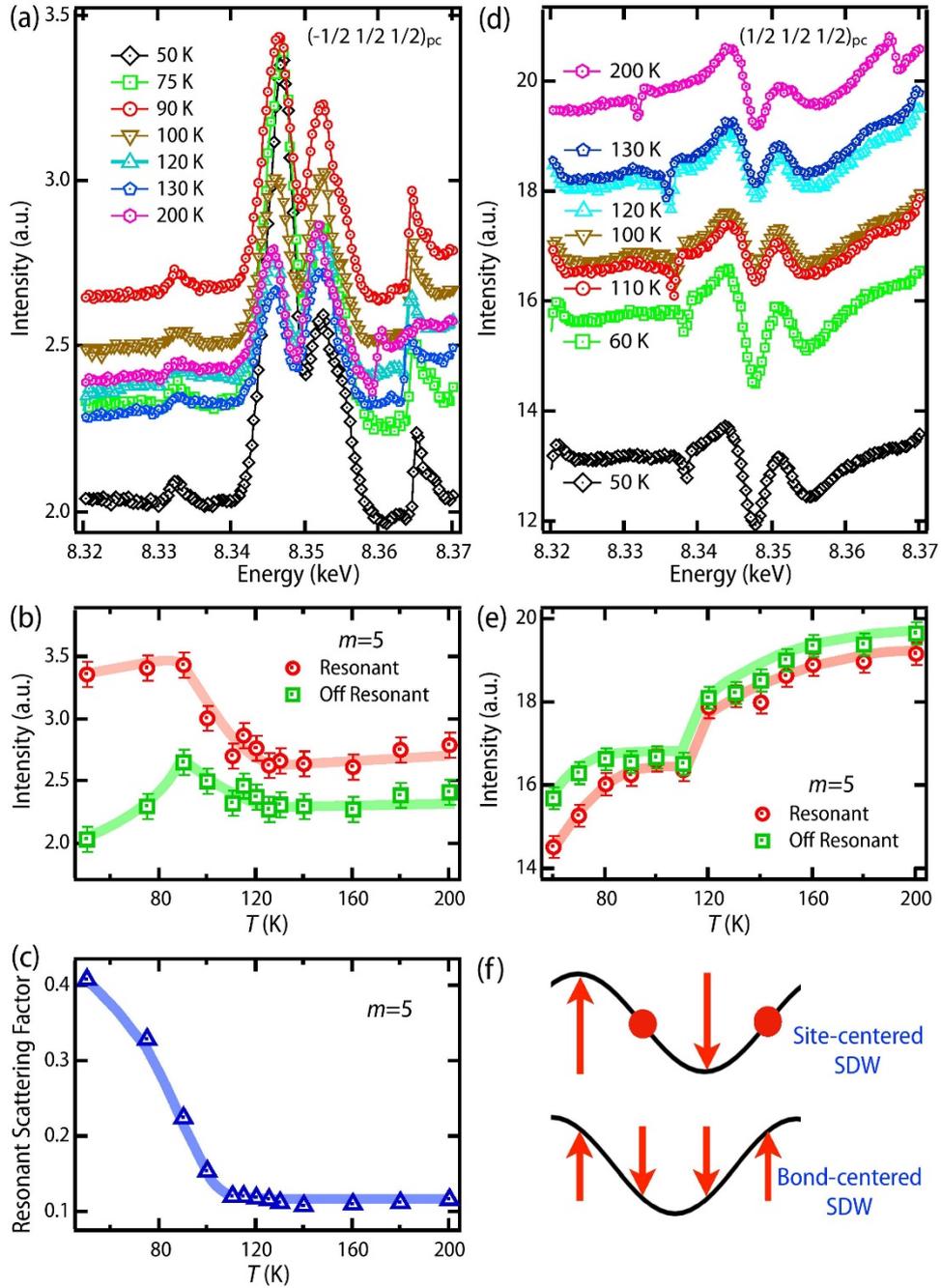

**Figure 4.** Resonance energy scan of the $m=5$ SL at various temperatures for (a) $(-1/2\ 1/2\ 1/2)_{pc}$ and (d) $(1/2\ 1/2\ 1/2)_{pc}$ Bragg peaks. Temperature dependence of the corresponding resonant intensity ($I_{res}$) and the off-resonant intensity ($I_{offres}$) are plotted in (b) and (e). (c) Variation of resonance enhancement factor ($\sqrt{I_{res}}-\sqrt{I_{offres}}$) for $(-1/2\ 1/2\ 1/2)_{pc}$ peak with $T$. (f) Site-centered SDW with charge ordering vs. bond-centered SDW without charge ordering.[46]

**Table 1.** Summary of transport data (heating run)

| m | $T_{MIT}$ (K) | $T_N$ (K) | NFL exponent (n) |
|---|---|---|---|
| 1 | 155 | 145 | 1 for T >180 K |
| 3 | 120 | 115 | 4/3 for 150 K < T <220 K; 1 for T >240 K |
| 5 | 120 | 110 | 1 for T >175 K |